\documentclass[showpacs,showkeys,11pt,
preprint,preprintnumbers,nofootinbib,
groupedaddress,superscriptaddress,amsmath,amssymb]{revtex4}
\usepackage{amsfonts}
\usepackage{amsmath}
\usepackage{graphics}
\usepackage{epsfig}
\usepackage{url}
\usepackage{multirow}
\usepackage{feynmp}

\newcommand {\be}{\begin{equation}}
\newcommand {\ee}{\end{equation}}
\newcommand {\ba}{\begin{eqnarray}}
\newcommand {\ea}{\end{eqnarray}}

\newcommand {\ra}{\rightarrow}

\begin{document}
\title{The $s$-channel Charged Higgs in the Fully Hadronic Final State at LHC}

\pacs{12.60.Fr, 
      14.80.Fd  
}
\keywords{Charged Higgs, MSSM, LHC}
\author{Ijaz Ahmed}
\email{Ijaz.ahmed@cern.ch}
\affiliation{National Center for Particle Physics, University of Malaya, 50603 Kuala Lumpur, Malaysia}
\affiliation{COMSATS Institute of Information Technology (CIIT), Islamabad 44000, Pakistan}
\author{Majid Hashemi}
\email{hashemi_mj@shirazu.ac.ir}
\affiliation{Physics Department and Biruni Observatory, College of Sciences, Shiraz University, Shiraz 71454, Iran}
\author{Wan Ahmad Tajuddin}
\email{wat@um.edu.my}
\affiliation{National Center for Particle Physics, University of Malaya, 50603 Kuala Lumpur, Malaysia}

\begin{abstract}
With the current measurements performed by CMS and ATLAS experiments, the light charged Higgs scenario  ($m_{H^{\pm}}$ $<$ 160 GeV), is excluded for most of the parameter space in the context of MSSM. However, there is still possibility to look for heavy charged Higgs boson particularly in the $s$-channel single top production process where the charged Higgs may appear as a heavy resonance state and decay to $t\bar{b}$. The production process under consideration in this paper is $pp \ra H^{\pm} \ra t\bar{b}~+~h.c.$, where the top quark decays to $W^{+}b$ and $W^{+}$ boson subsequently decays to two light jets. It is shown that despite the presence of large QCD and electroweak background events, the charged Higgs signal can be extracted and observed at a large area of MSSM parameter space ($m_{H^{\pm}}$,tan$\beta$) at LHC. The observability of charged Higgs is potentially demonstrated with 5$\sigma$ contours and $95\%$ confidence level exclusion curves at different integrated LHC luminosities assuming a nominal center of mass energy of $\sqrt{s}$ = 14 TeV.
\end{abstract}
\maketitle
\section{Introduction}
The neutral Standard Model (SM) Higgs boson with a mass of approximately 125 GeV was discovered by the CMS and ATLAS experiments \cite{Higgs_Discovery1,Higgs_Discovery2,Higgs_Discovery3} at CERN LHC in 2012 and marked a great triumph in the particle physics.
Most of the properties till now have been found consistent with those predicted for the SM Higgs boson.
However, the present scenario raises some interesting questions about the origin of the Electroweak Symmetry Breaking (EWSB).
It is undoubtedly said that the scalar sector of SM does engineer all of EWSB, but at the same time there are very convincing evidences from theoretical calculations and experimental signatures that SM needs to be superseded with other dynamics in order to consistently explain the issues regarding the dark matter in the universe, neutrino masses and naturalness problem.

Early attempts towards extending the SM scalar sector resulted in the Two Higgs Doublet Model (2HDM) \cite{2hdm1,2hdm2,2hdm3,2hdm4}, the Minimal Supersymmetric Standard Model (MSSM) \cite{mssm1,mssm2,mssm3} and Next to Minimal Sypersymmetric Standard Model (NMSSM) \cite{nmssm1,nmssm2}.

The discovery of another scalar boson, neutral or charged, would serve as unambiguous evidence for the new physics beyond the SM. The MSSM used as a benchmark in this paper is a special case of Type-II 2HDM. This model leads to five physical Higgs bosons: light and heavy CP-even Higgs bosons, h and H, a CP-odd Higgs boson, A, and two charged Higgs bosons, $H^{\pm}$. In this model, the couplings of the charged Higgs boson to up-type quarks are proportional to cot$\beta$ while the charged Higgs boson couplings to the down-type quarks and charged leptons are proportional to tan$\beta$, where tan$\beta$ is defined as the ratio of the vacuum expectation values of the two Higgs boson doublet fields.

The discovery of charged Higgs is quite challenging at particle colliders. On the other hand charged Higgs bosons provide unique signatures due to their electric charge which makes them different from neutral SM Higgs bosons in terms of their production, interaction and decay properties. Therefore there have been extensive searches for this particle over the last few years at Tevatron and LHC.

If the mass of charged Higgs $m_{H^{\pm}}$ is smaller than the mass difference between top and bottom quarks, $m_{H^{\pm}} < m_{t} - m_{b}$, the dominant production mechanism for the charged Higgs is via top quark decay: $t \rightarrow bH^{\pm}$. In this case the charged Higgs production is preferably produced via $t\bar{t}$ production process. Most of the studies performed at LEP, Tevatron and LHC focus on light charged Higgs mass domain, where charged Higgs predominantly decays into a pair of $\tau \nu$ $tan\beta > 5$ \cite{feynhiggs1,feynhiggs2,feynhiggs3} or into jets $(H^{\pm}\rightarrow cs)$.

In case of the heavy charged Higgs with $m_{H^{\pm}} > m_{t} + m_{b}$ the dominant production mode is the top quark associated production $H^{\pm}tb$. In this case, charged Higgs decay to a top quark, i.e., $H^{\pm}\rightarrow tb$, is kinematically allowed. However, identification of $t\bar{t}b\bar{b}$ signal in the presence of the huge irreducible background becomes difficult. Due to this reason, most early LHC analyses focus on the sub-dominant decay $H^{\pm}\rightarrow \tau \nu$ or $H^{\pm}\rightarrow cs$ in order to get advantage of suppressed backgrounds using $\tau$-identification tools.

Apart from $t\bar{t}$ production mechanism, the single top production processes at LHC have also been proved to be significant sources of charged Higgs in both low and high mass regions. Recently, there have been a number of analyses focusing on single top production as a source of charged Higgs. The light charged Higgs study has been performed in a t-channel single top production through top quark decay $(pp\rightarrow tq\rightarrow qbH^{\pm}\rightarrow qb\tau\nu)$ if the $\tau$ lepton decays hadronically \cite{st1} or leptonically \cite{SingleTop_CH}.
The heavy charged Higgs has been analyzed through s-channel single top production in the leptonic final state $(pp\rightarrow tb\rightarrow bbW^{\pm}\rightarrow bbl^{\pm}\nu_{l})$ \cite{st2}. The off-diagonal couplings between incoming quarks in the s-channel single top production have also been studied leading to an enhancement of the total cross-section by a factor of 2.7 \cite{myplb}. Similarly in \cite{st3} the t-channel single top production has been considered as a source of charged Higgs exchange, though being observable at very high integrated luminosities and high tan$\beta$ values.

In \cite{atlas_singletop_CH} and \cite{cdf_singletop_CH}, the $s$-channel single top has been considered as a source of charged Higgs production and decay to $t\bar{b}$ where the $W$ boson from the top quark decay, undergoes a hadronic decay to a pair of light jets. To the best of our knowledge, no more detailed analysis of this type exists in the literature. The aim of this paper is to study the s$-$channel single top in the chain $pp\rightarrow H^{\pm}\rightarrow tb \rightarrow bbW \rightarrow bbj_{1}j_{2}$ at LHC using new techniques and generators focusing on the charged Higgs mass in the available area of the parameter space which has not yet been excluded by LHC data, i.e., $200 ~<~m_{H^{\pm}}~<~400$ GeV. There are background processes like QCD multi jets and W+jets which make it a challenging analysis. However, as will be seen, they can be well under control.

In the following sections, signal and background events are introduced and their cross sections are presented. An event selection and analysis is described in detail with the aim of charged Higgs invariant mass reconstruction with different mass hypotheses. Finally an estimation of accessible regions of MSSM parameter space ($m(H^{\pm}$),$tan\beta$) for a 5$\sigma$ discovery or exclusion at 95$\%$ C.L. is provided. The theoretical framework is based on MSSM, $m_h-max$ scenario with the following parameters: $M_{2}$ = 200 GeV,~$M_{\tilde{g}}$ = 800 GeV,~$\mu$ = 200 GeV and $M_{SUSY}$ = 1 TeV. The $m_h-max$ scenario defines a benchmark point optimized to maximize the theoretical upper bound on $m_{h}$ for a given tan$\beta$ and fixed $m_{t}$ and the soft SUSY breaking parameter $M_{SUSY}$. This benchmark point provides the largest parameter space in the $m_{h}$ direction and conservative exclusion limits for tan$\beta$.

\section{Experimental Constraints}
\subsection{Direct Searches}
The charged Higgs search has been performed for decades, at colliders like LEP  \cite{lepexclusion1}, Tevatron \cite{cdf3} and LHC using the ATLAS \cite{CHtaunu8TeVATLAS,CHIndirect8TeVATLAS,LightCH7TeVATLAS2,LightCH7TeVATLAS1} and CMS \cite{cms2,cms3} experiments.
Experimental exclusion limits are set by the LEP experiments at $m_{H^{\pm}}$ $>$ 79.3 GeV at 95$\%$ C.L. independently of the branching ratios \cite{LEPDirect}, assuming $BR(c\bar{s})+BR(\tau\nu)$ = 1. Assuming $BR(\tau\nu)$ = 1 for the low charged Higgs mass, the set limit is 87.8 GeV. For the heavier charged Higgs boson the best current limits are set by ATLAS and CMS.

For the charged Higgs mass range 80 GeV $< m_{H^{\pm}} <$ 160 GeV, ATLAS imposes 95$\%$ CL upper limits on $BR(t\rightarrow H^{\pm}b$) in the range $0.23-1.3\%$, and for the mass range 180 GeV $< m_{H^{\pm}} <$ 1000 GeV, 95$\%$ CL upper limits on the production cross-section in the range $0.0045-0.76$ pb, both with assumption that $BR(H^{\pm}\rightarrow \tau\nu)=1$ \cite{LightCH7TeVATLAS1}.

Similarly in the mass range 80 GeV $<$ $m_{H^{\pm}}$ $<$ 160 GeV, 95$\%$ CL upper limits on $BR(t\rightarrow H^{\pm}b)$ are set in the range 0.16 - 1.2$\%$, and for 180 GeV $<$ $m_{H^{\pm}}$ $<$ 600 GeV, 95$\%$ CL upper limits on production cross-section of charged Higgs are set in the range $0.026-0.38$ pb, both with assumption that $BR(H^{\pm}\rightarrow \tau\nu)$ = 1  in CMS Experiment \cite{cms3}.
\subsection{Indirect searches}
The exclusion limits from the indirect searches can be obtained by studying flavor physics, measuring electric dipole moment of electron or other precision measurements \cite{CHLHC_CPViolating2HDM}. These limits are highly model dependent and can not replace the direct searches. On the other hand, they are generally for 2HDM and translating them to the case of a supersymmetric model like MSSM is not trivial.
\begin{enumerate}
\item In $b\rightarrow s\gamma$ decay, a charged Higgs boson can contribute and change the branching fraction with respect to the SM-only scenario and can therefore be used to probe physics beyond the SM. In \cite{B2sgamma}, indirect mass constraints at 95$\%$ C.L. are set via $B\rightarrow X_{s}\gamma$, excluding charged Higgs boson in the 2HDM type II up to 295 GeV.
\item There are other processes where further constraints can be used in indirect searches e.g., $B_{u}\rightarrow \tau\nu_{\tau}$ \cite{BU2taunu1,BU2taunu2}, $B\rightarrow D\tau\nu_{\tau}$ \cite{Ds2taunu}, $D_{s}\rightarrow\tau\nu_{\tau}$ \cite{D2taunu} and $B_{d,s}\rightarrow \mu^{+}\mu^{-}$ \cite{B2mumu}.
\end{enumerate}

\begin{figure}[htb]
 \begin{center}
 \includegraphics[width=0.6\textwidth]{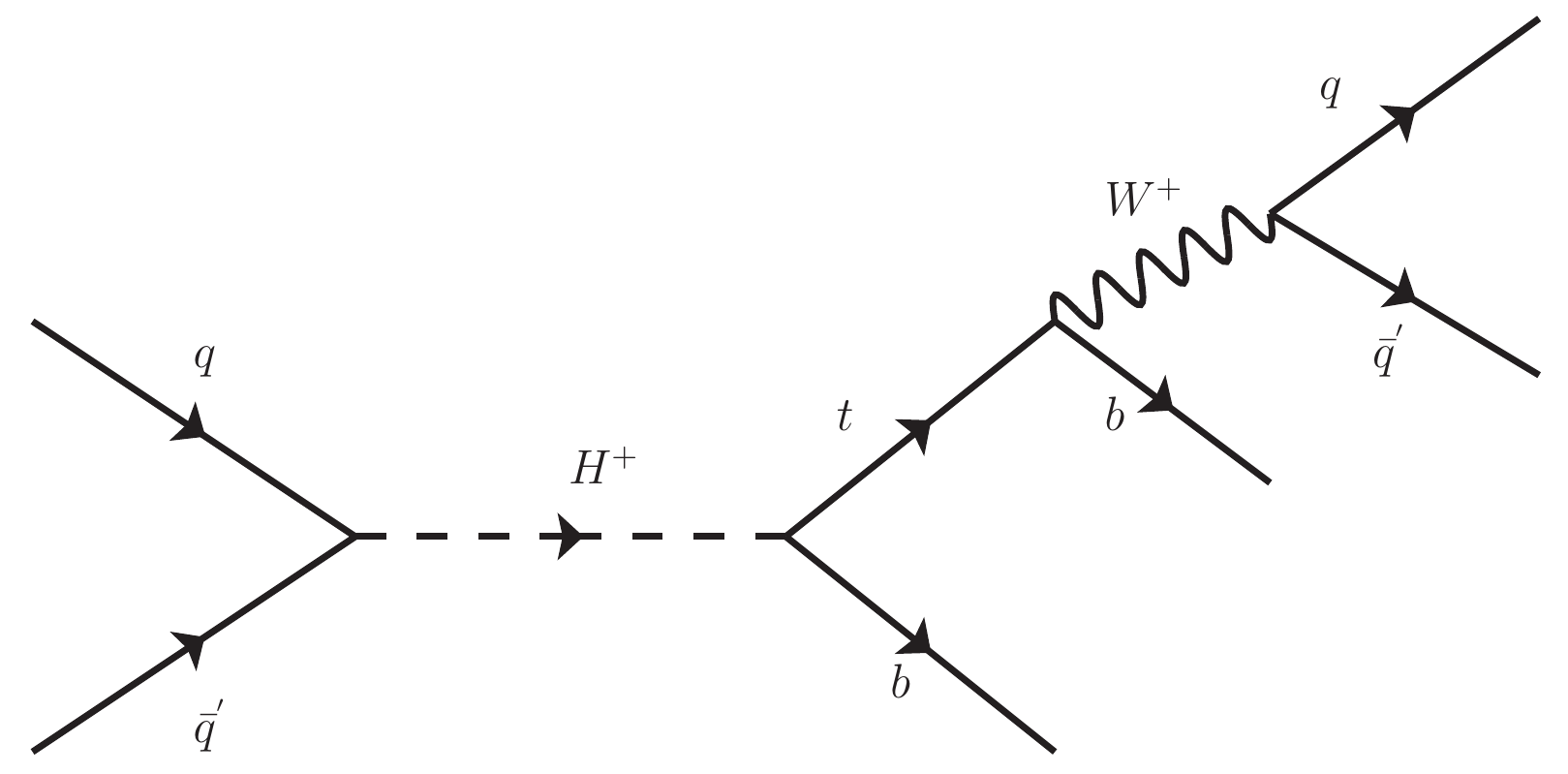}
 \end{center}
 \caption{The s-channel single top production diagram as a signal process with its full hadronic decay mode.}
 \label{sig1}
 \end{figure}

\section{Signal and background processes and their cross-sections}

The single top production occurs through electroweak interactions and proceeds through three different processes at the LHC depending on the virtuality of the W-boson involved in SM. In the t-channel, the W-boson is space-like ($q^{2}_W \le 0$). This process is the largest source of single top production in SM and its cross-section is around one third of the $t\bar{t}$ cross-section. The signature of this channel is a high momentum forward light quark and a single top quark. In the s-channel process, the involved W-boson is time-like ($q^{2}_{W} \ge 0$). In this case a top quark and a hard b-quark are produced in final state. In the associated $tW$ production channel, the W-boson is real ($q^{2}_{W} = m^{2}_{W}$). This process involves a b-quark and a gluon both from proton sea, a real W-boson and a top quark. In MSSM the single top production processes have the same final state but with the possibility of exchanging the charged Higgs boson which changes the kinematics of the final state particles.

The charged Higgs boson has been considered as a crucial signature of MSSM at LHC due to its distinct nature of decay channels. The search for heavy and light charged Higgs are quite different due to their different decay modes and topological parameters. In the s-channel single top production, a charged Higgs is produced in the intermediate phase as a heavy state decaying to a top and b quark in the five flavor scheme $(H^{\pm}\rightarrow t\bar{b}$) as shown in Fig.\ref{sig1}.
However in four flavor scheme a top quark, a b quark and a light quark are produced in the final state. The top quark exclusively decays into a b quark and W-boson because of the largest coupling between top and b quark which leads to $V_{tb}\approx 1$, while the W-boson in the top quark decay undergoes a hadronic decay, i.e., two jets in the final state.
Events in which the top quark decays into a fully hadronic final state are interesting from different aspects. They constitute the largest branching fraction $\approx 68\%$ related to the W boson decay and thus the signal statistics is larger than the leptonic final state. There is also possibility to use them in mass reconstructions for the determination of the top-quark and charged Higgs mass. This is the reason of using the hadronic final state in this analysis.

\begin{figure}[htb]
 \begin{center}
 \includegraphics[width=0.6\textwidth,natwidth=610,natheight=642]{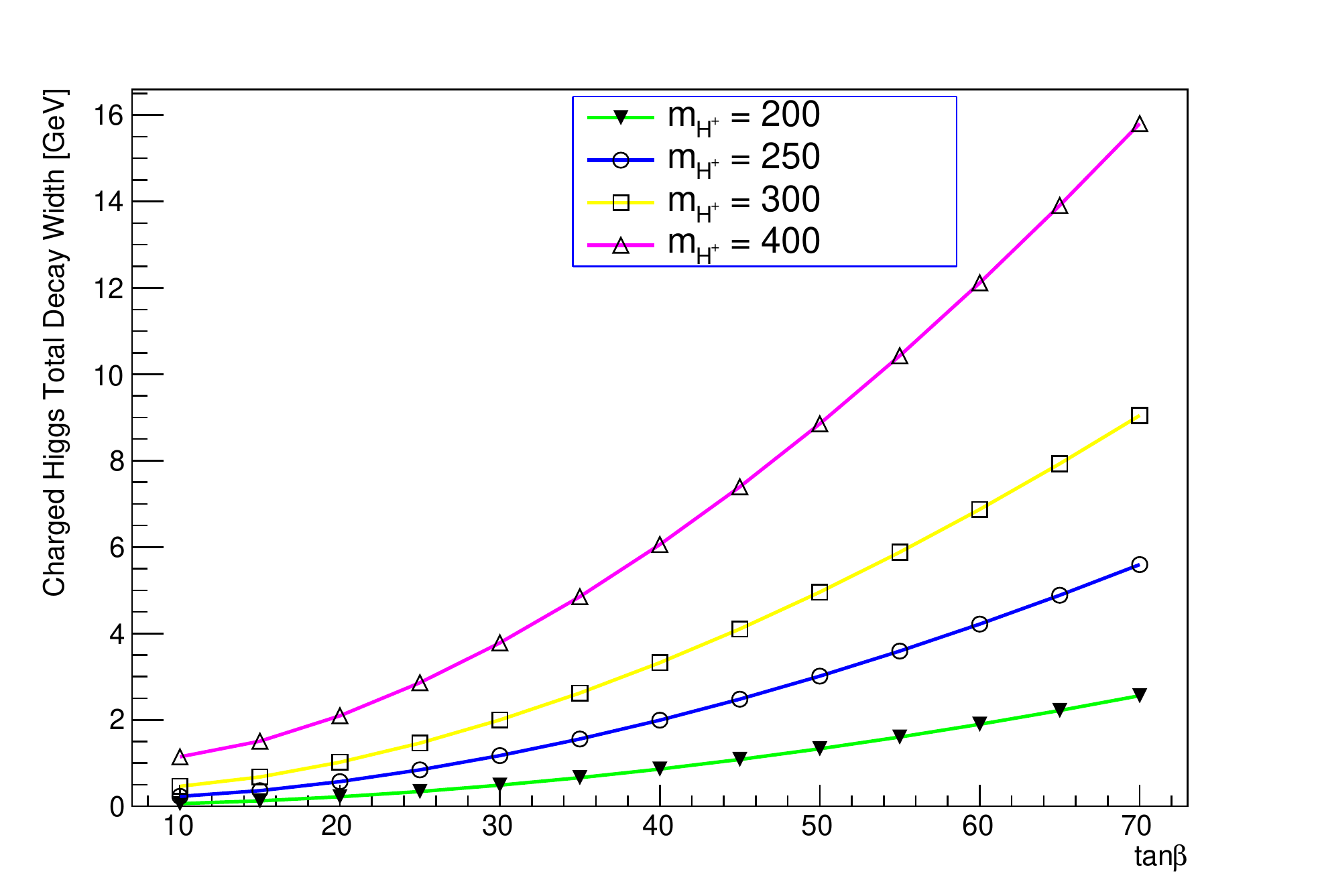}
 \end{center}
 \caption{The total decay width of charged Higgs at various Higgs mass values are shown as a function of tan$\beta$}
 \label{width}
 \end{figure}

The main background processes are $W^{\pm}jj$, $W^{\pm}b\bar{b}$, $W^{\pm}c\bar{c}$, $t\bar{t}$, s-channel and t-channel single top in SM and QCD multijets production. The cross-section of all backgrounds are computed and samples are generated by PYTHIA 8.1.53 \cite{pythia} except $W^{\pm}jj$, $W^{\pm}b\bar{b}$ and $W^{\pm}c\bar{c}$ (also called "W+2jets") which are calculated using Madgraph \cite{madgraph1,madgraph2} with a kinematic preselection cut applied as $P^{jets}_{T}>20~$GeV. The W+jets events from Madgraph and also signal events generated using CompHEP \cite{comphep1,comphep2} are obtained as output files in the LHA format \cite{lhef} and passed to PYTHIA for multi-particle interaction, parton showering and hadronization. The corresponding cross-sections of all these process are listed in the Table 2. For signal cross-section calculation CompHEP package is used using the charged Higgs total decay width calculated by FeynHiggs \cite{feynhiggs1,feynhiggs2,feynhiggs3}. The results for decay widths are shown in Fig.\ref{width}.

\begin{figure}[tb]
 \centering
   \includegraphics[width=0.6\textwidth]{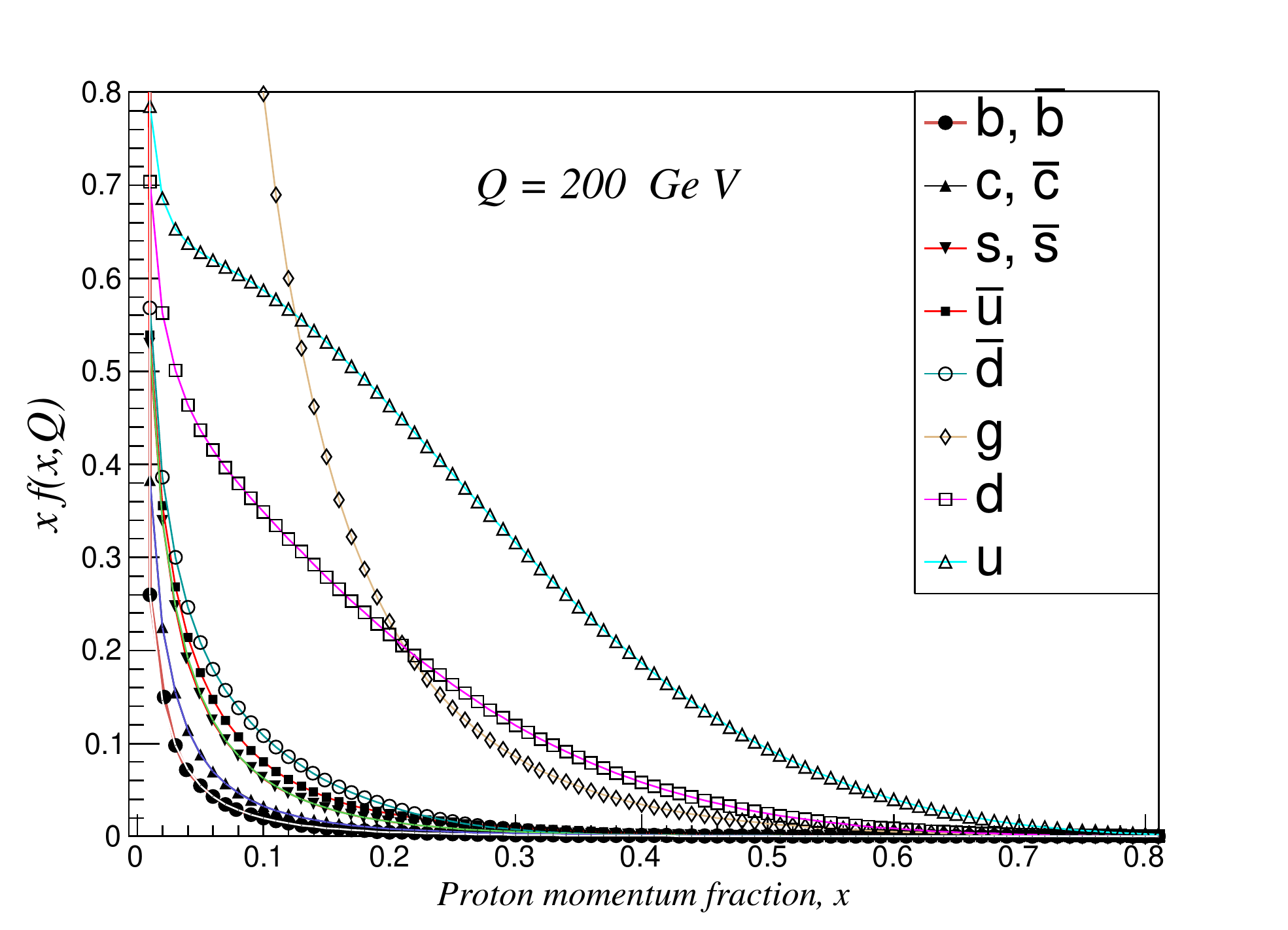}
   \caption{Parton distribution functions at the scale of $Q=200$ GeV.}
   \label{pdf}
 \end{figure}

The cross section of the signal includes both diagonal and off-diagonal contributions of the incoming partons. In order to see the relative contribution of each incoming parton pair, the cross section formula can be written in terms of the product of the charged Higgs partial decay rates, $\Gamma(H^{\pm}\to UD)\Gamma(H^{\pm}\to U'D')$ for a signal process as $UD \to H^{\pm} \to U'D'$. Here $UD$ ($U'D'$) is incoming (outgoing) parton pair. This is, however, a partonic interaction which should be convoluted with parton distribution functions $f(x,Q,i)$ where $x$ is the proton momentum fraction carried by the parton, $Q$ is the momentum transfer (set to the charged Higgs mass) and $i$ is the parton index. Fig.\ref{pdf} shows these functions for $Q=200$ GeV. An integration over all $x$ values from zero to unity gives the total cross section including all possible contributions. 

\begin{figure}[tb]
 \centering
   \includegraphics[width=0.6\textwidth]{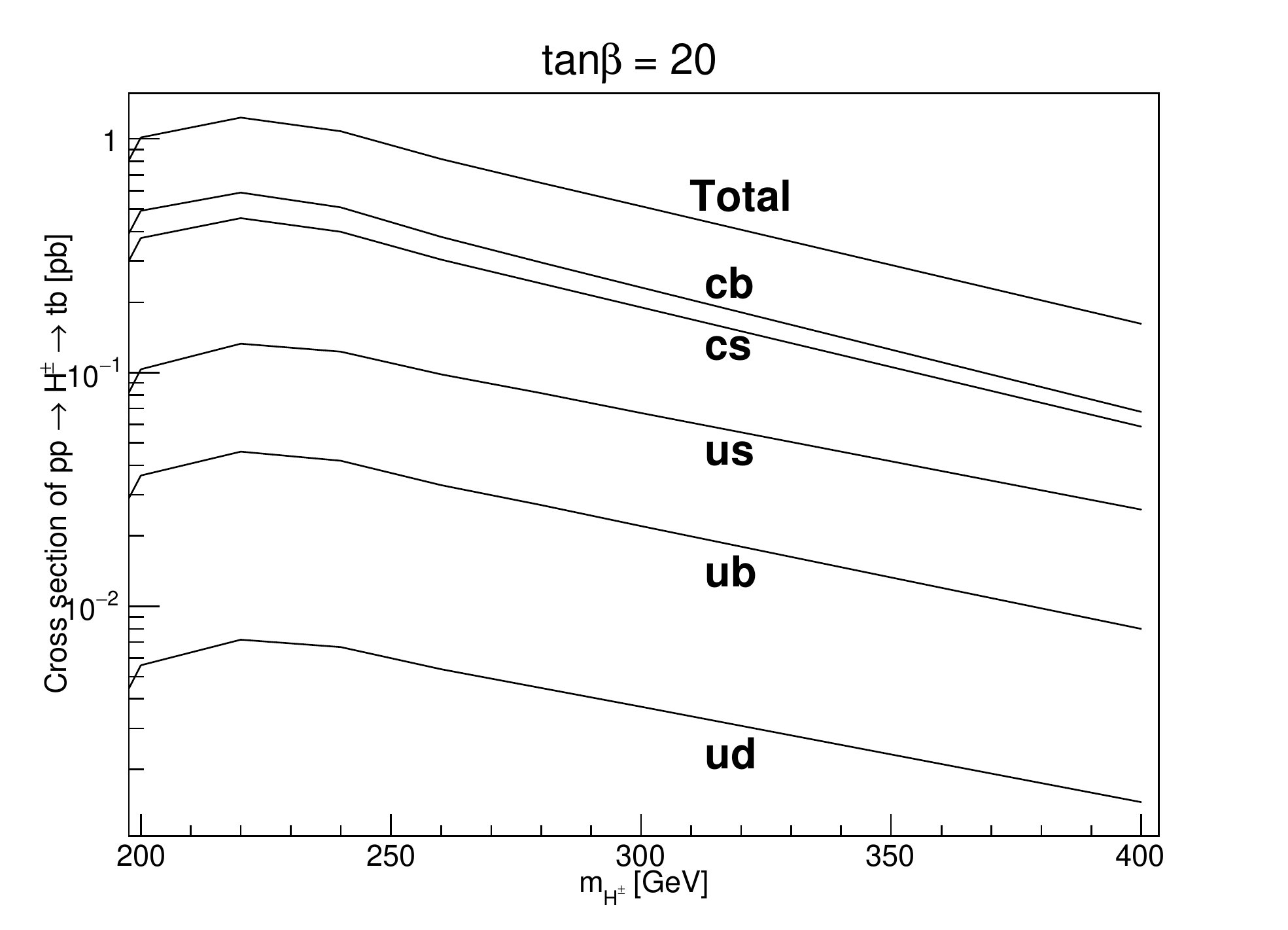}
   \caption{Cross section of different contribution of parton pairs as a function of the charged Higgs mass at $\tan\beta=20$.}
   \label{cs}
 \end{figure}

It may not be obvious how different contributions are compared but the $c\bar{b}$ incoming pair has the largest contribution to the cross section. In fact its contribution is larger than the diagonal contribution of $c\bar{s}$ pair. The reason is as follows. The charged Higgs decay rate is proportional to the square of the CKM matrix element as well as the square of the down type quark mass at high $\tan\beta$ values (for $\tan\beta$ values considered in this paper, this is a good approximation). This is shown in Eq. 1.
\begin{equation}
\Gamma_{H^{\pm}\to UD}=\frac{3\sqrt{2}G_{F}{V_{UD}}^2}{8\pi}m_{H^{\pm}}\left( 1-\frac{m_{U}^2}{m_{H^{\pm}}^2} \right) [m_{U}^2\cot^2\beta+m_{D}^2\tan^2\beta]
\end{equation}
A comparison of the parton distribution functions (shown in Fig. \ref{pdf}) shows that the $b$-quark distribution is roughly one third of that of the $c$-quark. The ratio of cross sections of the two incoming states can be written as 
\begin{equation}
\frac{\hat{\sigma}_{c\bar{b}}}{\hat{\sigma}_{c\bar{s}}}=\frac{\Gamma_{c\bar{b}}}{\Gamma_{c\bar{s}}}=\frac{V_{c\bar{b}}^2~m_b^2}{V_{c\bar{b}}^2~m_s^2}.
\end{equation}     
Therefore the ratio of differential cross sections for a given $x$ and $Q$ is 
\begin{equation}
\frac{d\sigma_{c\bar{b}}}{d\sigma_{c\bar{s}}}=\frac{\hat{\sigma}_{c\bar{b}}~f(x,Q,b)}{\hat{\sigma}_{c\bar{s}}~f(x,Q,s)}=\frac{V_{c\bar{b}}^2~m_b^2~f(x,Q,b)}{V_{c\bar{b}}^2~m_s^2~f(x,Q,s)}.
\end{equation}     
Using quark masses at the scale of $Q = 200$ GeV, i.e., $m_{b} = 2.63$ GeV and $m_{s} = 0.05$ GeV, and the CKM matrix elements as $V_{c\bar{b}} = 0.04$ and $V_{c\bar{s}} = $1, and the ratio of parton distribution functions equal to 3, one would obtain the ratio of cross sections to be $\sim 1.5$. This factor means that a large part of the total cross section comes from the off-diagonal contribution of $c\bar{b}$ pair. Therefore the total cross section is $\sim 2.5$ times that of the $c\bar{s}$ initiated process. Fig. \ref{cs} shows contribution of each incoming state to the total cross section.
 
\begin{figure}[htb]
 \begin{center}
 \includegraphics[width=0.6\textwidth,natwidth=610,natheight=642]{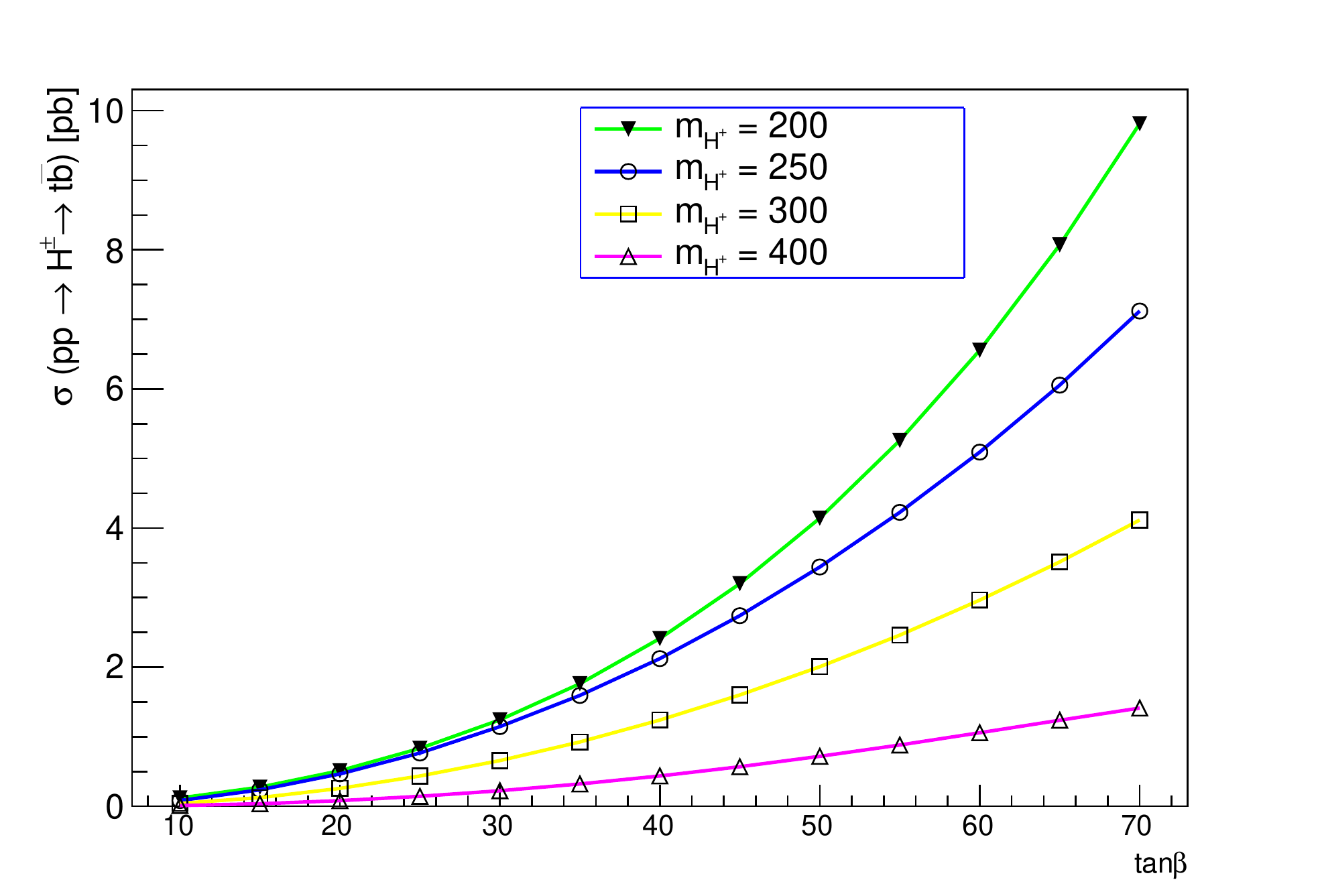}
 \end{center}
 \caption{The s-channel charged Higgs production cross-section as a function of tan$\beta$ and charged Higgs masses.}
 \label{sigma}
 \end{figure}

The integration over parton level cross-sections is performed using CTEQ 6.6 parton distribution function (PDF) provided by LHAPDF 5.9.1 \cite{lhapdf} at nominal LHC centre of mass energy $\sqrt{s}$= 14 TeV. The total cross-section is the sum of all initial states, i.e., the diagonal and off-diagonal couplings as shown in Fig. \ref{sigma} at various $tan\beta$ values. In order to get the $\sigma \times BR(W^{+}\rightarrow jj)$, a factor of 0.68 is multiplied to all the signal and background cross-sections to ensure fully hadronic final state.  Jet reconstruction is performed with the FASTJET 3.1.3 \cite{fastjet1,fastjet2} using anti-kt algorithm \cite{antikt} and $E_T$ recombination scheme with a cone size of $\Delta R$=0.4, where $\Delta R=\sqrt{(\Delta \eta)^{2}+(\Delta \phi)^2}$ with $\eta=-\ln\tan(\theta/2)$ and $\theta(\phi)$ are the polar (azimuthal) angles with respect to the beam pipe defined as z-axis.

\section{Event Selection and Analysis}
The approach used in this analysis is the same as a typical physics channel analysis in the sense that first the signal and corresponding background samples having similar final states are identified and then cross sections are calculated by the event generators. Decay widths of particles can be calculated using available packages and used in cross section calculation. The algorithm then starts with optimized selection cuts using kinematic features of the signal and background and their kinematic differences and eventually the signal statistical significance after all selection cuts is calculated. Different mass windows may be applied at each point for selecting W boson, top quark and charged Higgs invariant mass distributions. The detailed analysis is expressed as follows.

The final state of signal events as shown in Fig. \ref{sig1} contains a collection of 2 lights jets and 2 $b$-jets. First the jets reconstruction is performed and jets are selected if they satisfy the requirement of having $E^{jet}_{T}>20$ GeV and $|\eta|<2.5$, where $E_{T}$ is transverse energy of the jet and $\eta$ is pseudo-rapidity defined previously. An event has to have 4 jets passing above requirement, two of which are b-tagged.

The b-tagging is emulated by a jet-quark matching algorithm which calculates the spatial distance between the reconstructed jet and a $b$ or $c$ quark from generator level information in terms of $\Delta R$. If $\Delta R (jet,quark)~<~0.2$ with $p_{T}>50~GeV$ and $|\eta|<3.0$, the jet is flagged as a $b$-jet. The $b$-jet efficiency is assumed to be $60\%$ while $c$-jet mis-tagging rate is taken to be $10\%$. The existence of 2 b-jets in the event is expected to dramatically suppress the Wjj and QCD sample. However, as will be seen that the Wjj and $t\bar{t}$ events are the main background.

\begin{figure}[htb]
 \begin{center}
 \includegraphics[width=0.6\textwidth,natwidth=610,natheight=642]{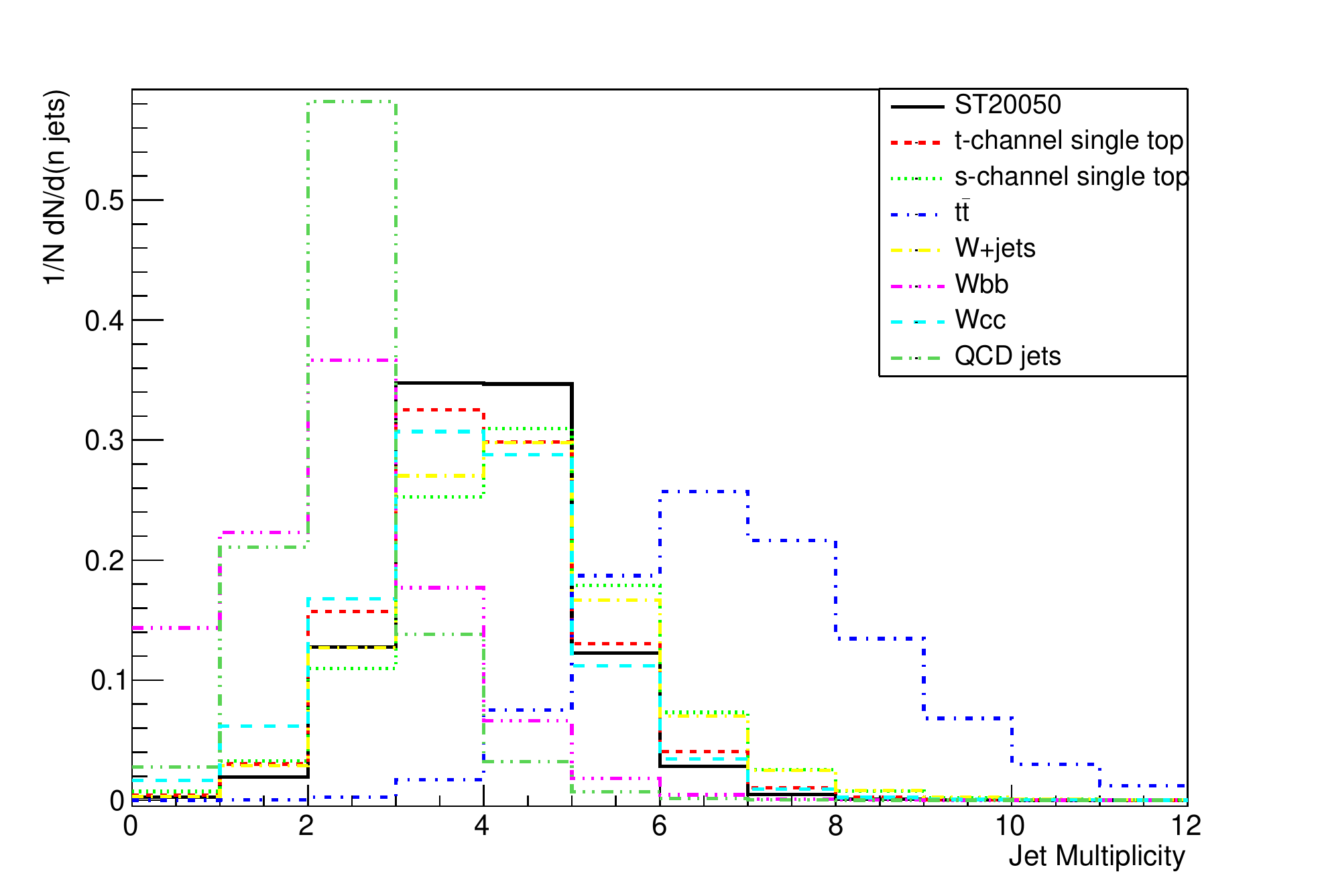}
 \end{center}
 \caption{The jet multiplicity distribution is shown with both signal and background events.}
 \label{jetmul}
 \end{figure}

\begin{figure}[h]
 \begin{center}
 \includegraphics[width=0.6\textwidth,natwidth=610,natheight=642]{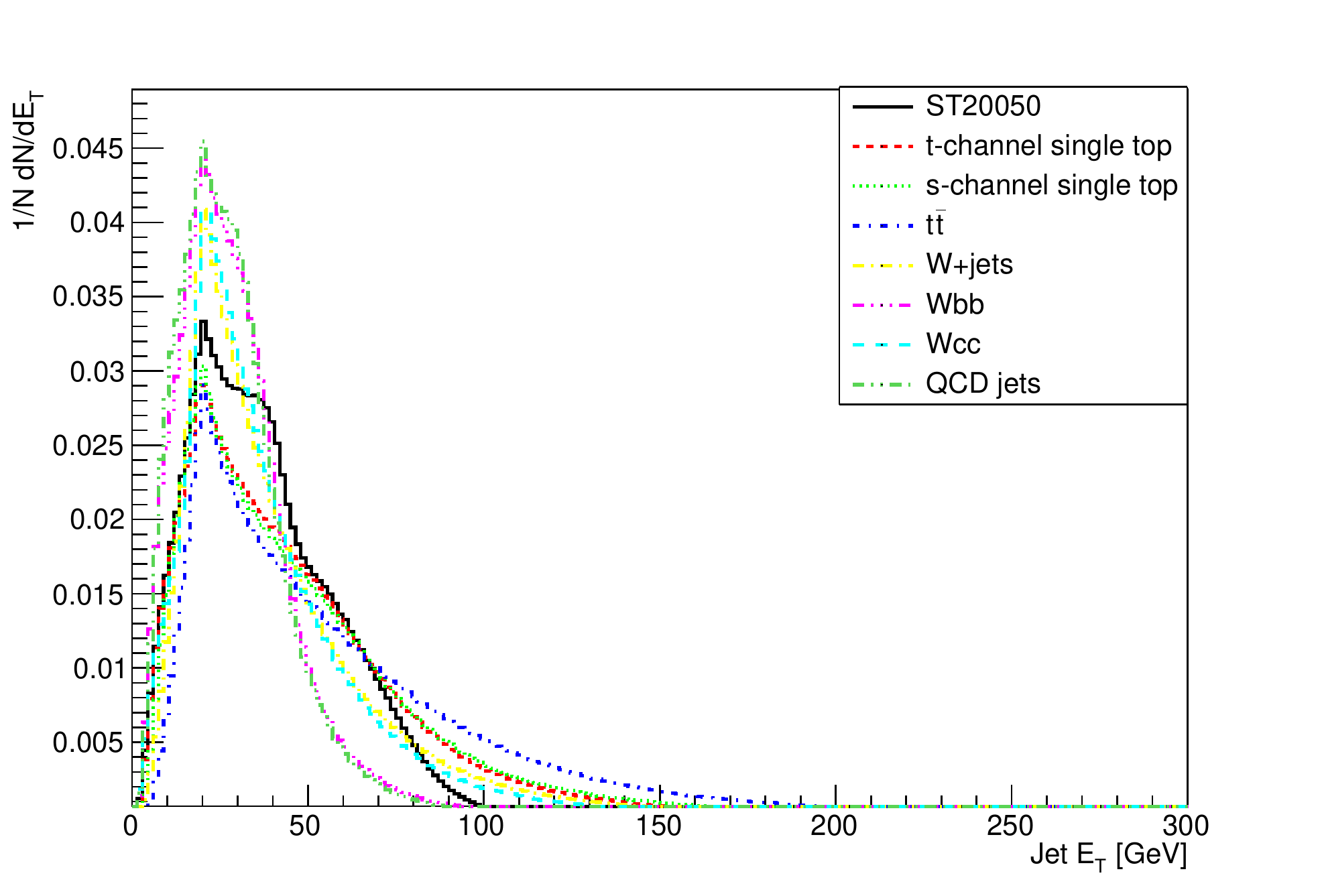}
 \end{center}
 \caption{The reconstructed jets transverse energy distribution.}
 \label{jetet}
 \end{figure}

\begin{figure}[h]
 \begin{center}
 \includegraphics[width=0.6\textwidth,natwidth=610,natheight=642]{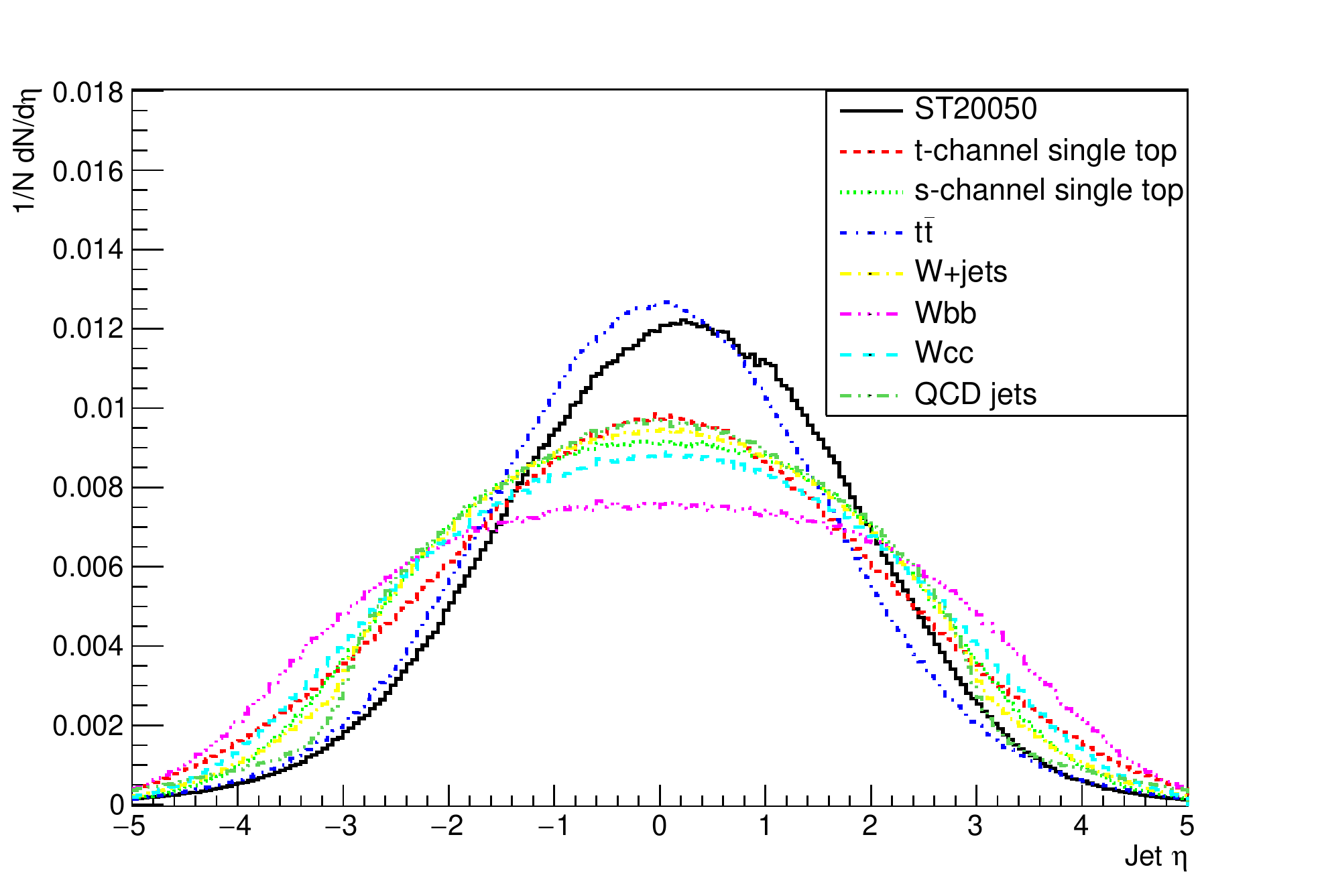}
 \end{center}
 \caption{Pseudorapidity distribution of selected jets}
 \label{jeteta}
 \end{figure}

\begin{figure}[htb]
 \begin{center}
 \includegraphics[width=0.6\textwidth,natwidth=610,natheight=642]{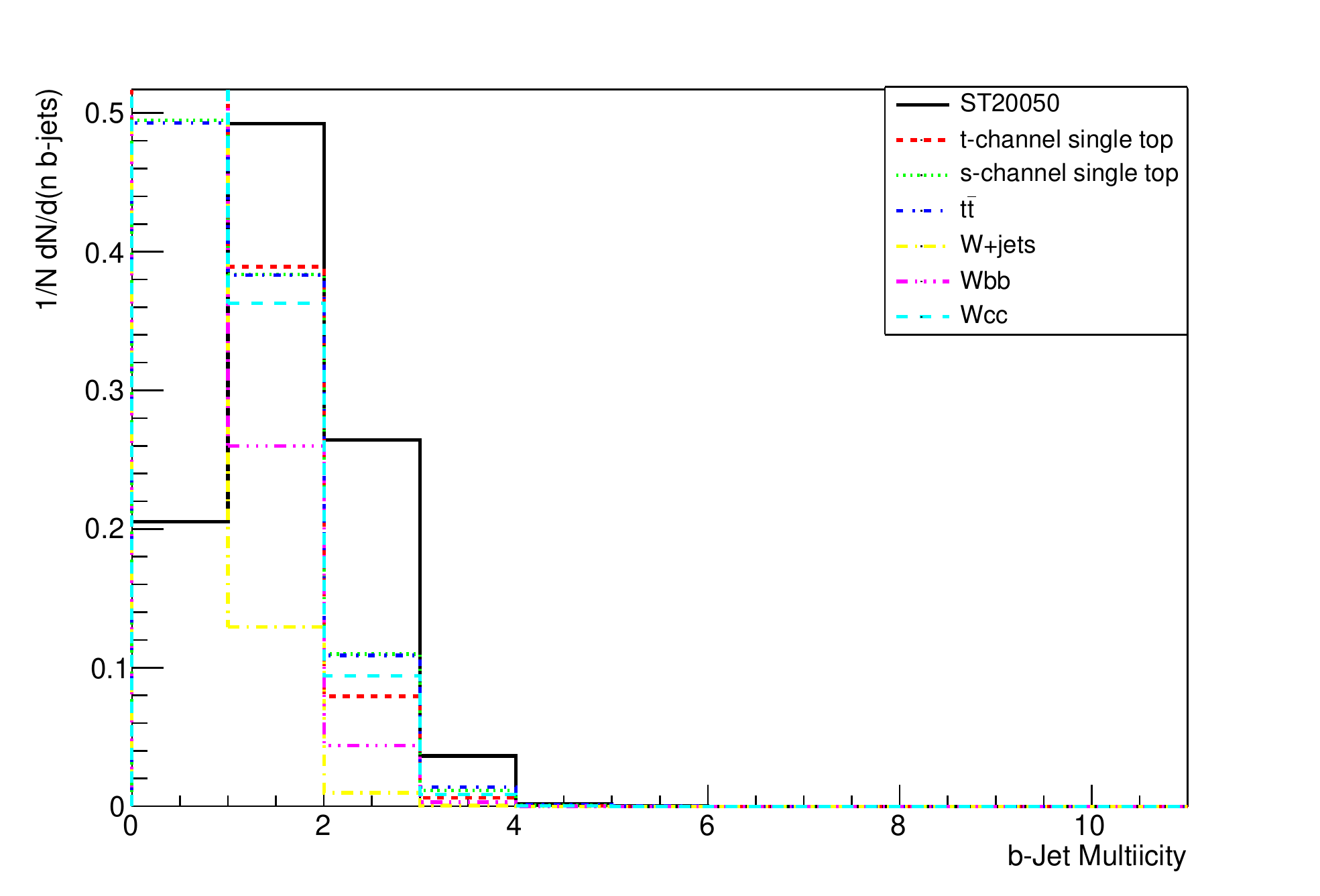}
 \end{center}
 \caption{The b-jet multiplicity distribution in both signal and background events.}
 \label{bjetmul}
 \end{figure}

\begin{figure}[htb]
 \begin{center}
 \includegraphics[width=0.6\textwidth,natwidth=610,natheight=642]{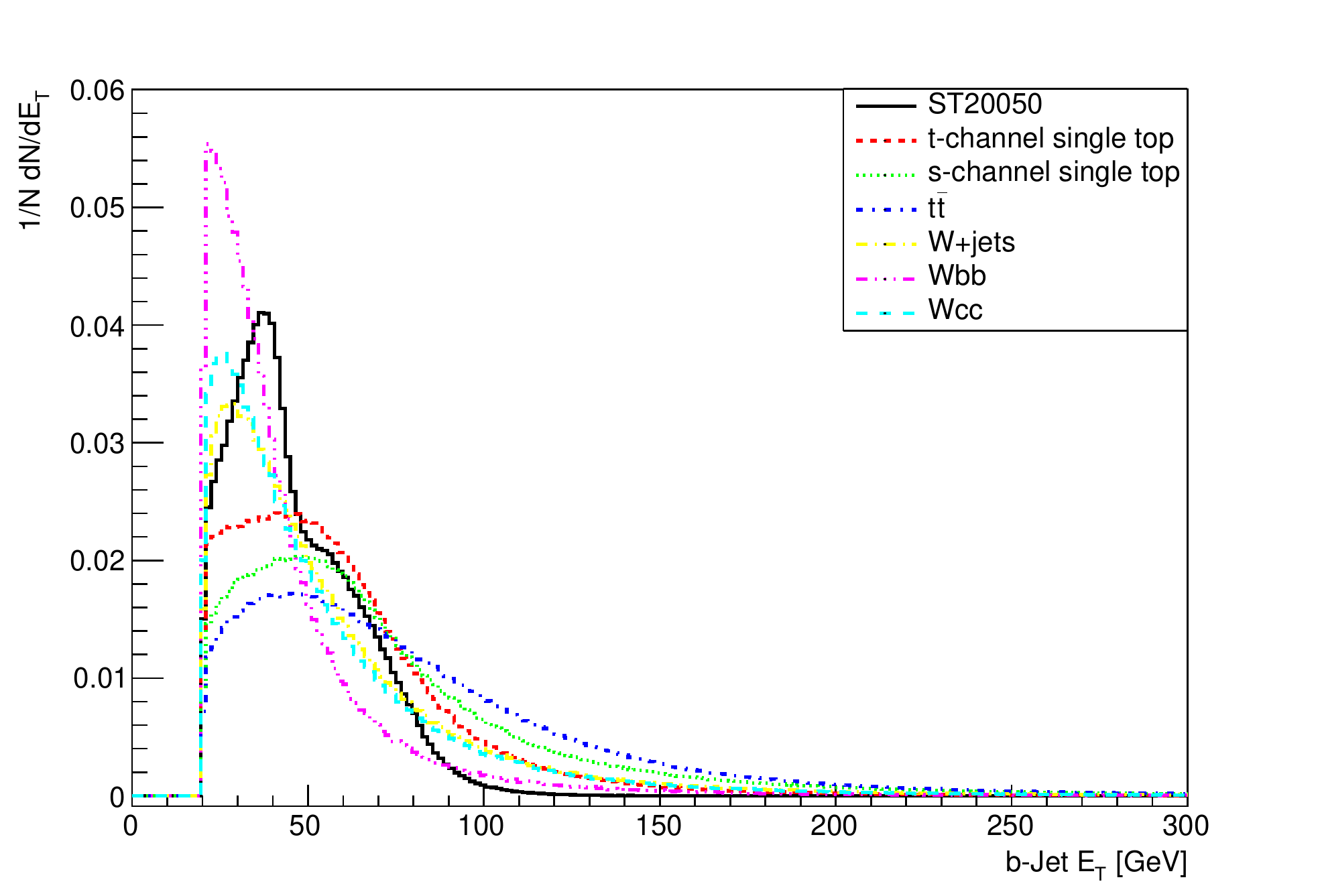}
 \end{center}
 \caption{The b-jet transverse energy distribution is shown in signal and background events.}
 \label{bjetet}
 \end{figure}

The jets which do not satisfy the $b$-jet requirement are declared as light jets. For W boson invariant mass reconstruction, two leading jets are selected with same $p_{T}$ and $\eta$ cuts applied on all jets. The low jet multiplicity is a feature of signal events which can be used to suppress $t\bar{t}$ events
and single W events accompanied by more than two jets. Fig. \ref{jetmul} shows a comparison between signal and background events in terms of their jet multiplicities.  Throughout the paper, plots are shown with a signal comprising of a charged Higgs mass $m_H^{\pm}$ = 200 GeV and $\tan\beta$ = 50 abbreviated as "ST20050". The $tan\beta$ factor only contributes to the signal cross-section without changing the shape of distributions. In Fig. \ref{jetet} and Fig. \ref{jeteta}, the light jet transverse momentum and pseudo-rapidity distributions are plotted for signal and background events. Similar to light jets, Figs. \ref{bjetmul} and \ref{bjetet} show b-jets multiplicity comparisons in both signal and background and transverse energy distributions respectively.

\begin{figure}[h]
 \begin{center}
 \includegraphics[width=0.6\textwidth,natwidth=610,natheight=642]{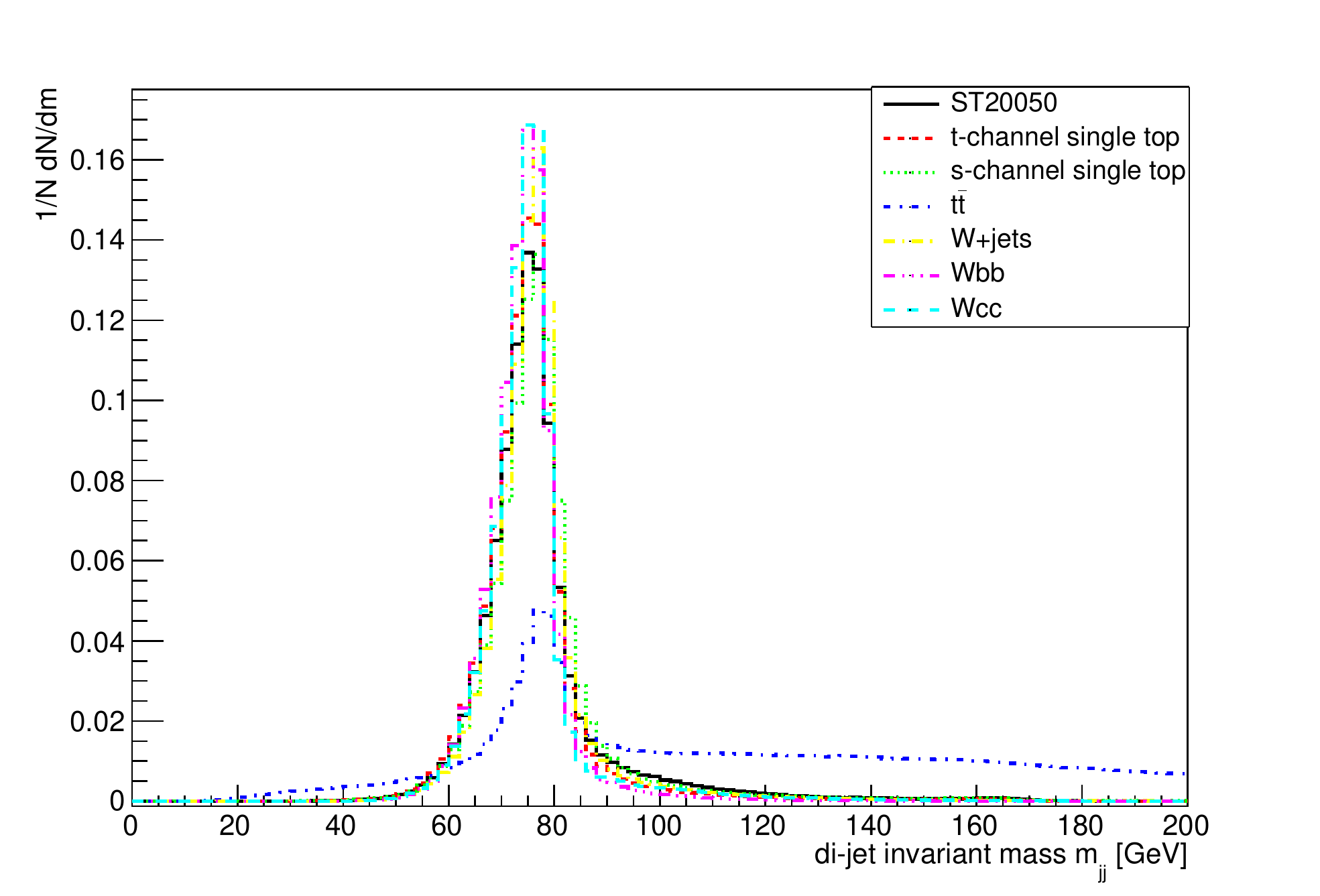}
 \end{center}
 \caption{The reconstructed W-boson invariant mass in signal and events background}
 \label{wmasshad}
 \end{figure}

\begin{figure}[h]
 \begin{center}
 \includegraphics[width=0.6\textwidth,natwidth=610,natheight=642]{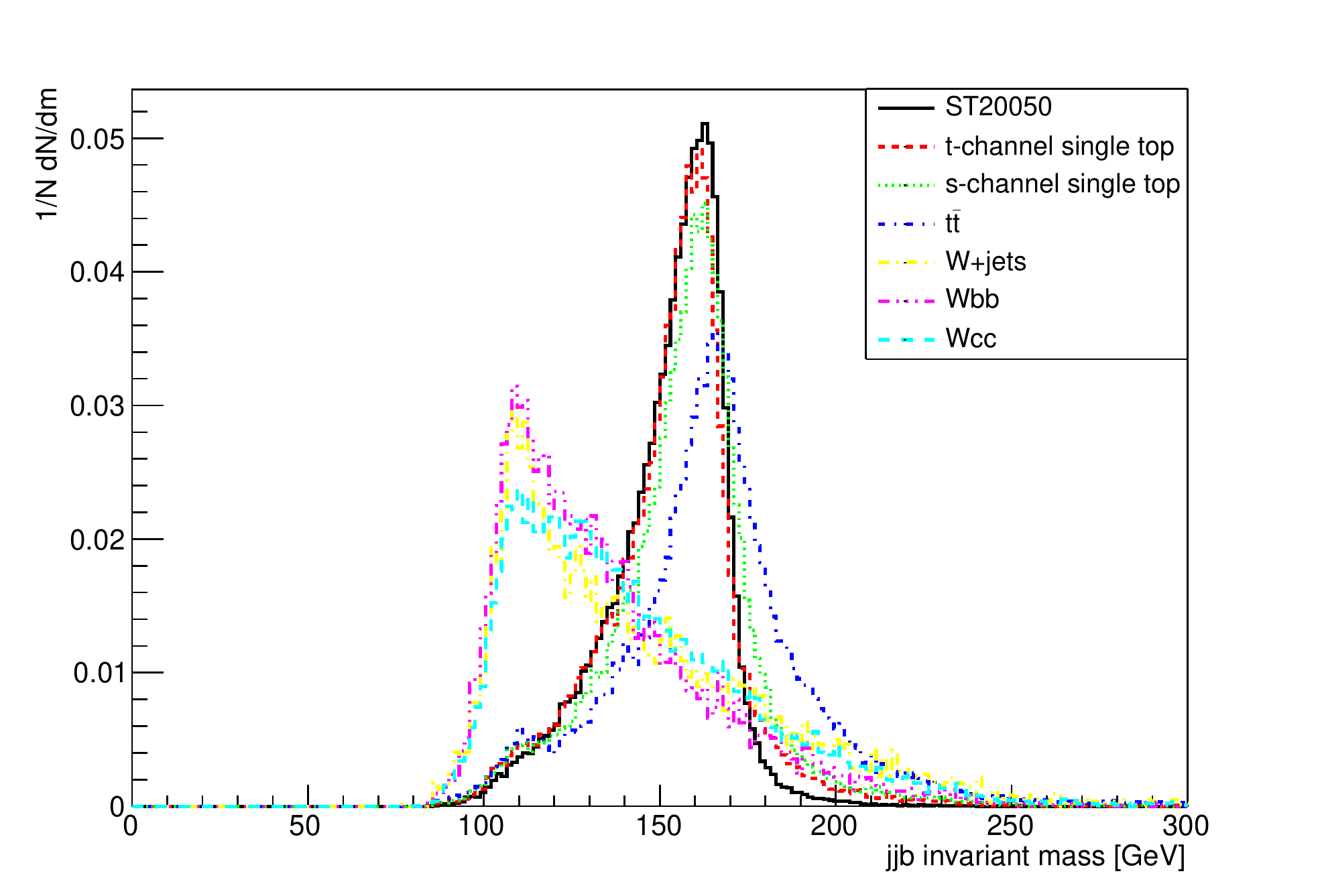}
 \end{center}
 \caption{The reconstructed invariant mass distribution of $m_{jjb}$ signal and dominant backgrounds. }
 \label{topmass}
 \end{figure}

The two highest $p_{T}$ light jets are combined together to form the W-boson candidate as plotted in Fig. \ref {wmasshad}. The top quark candidate invariant mass distribution is obtained by combining three jets, i.e., two light jets and a $b$-jet which gives the closest top quark invariant mass to its nominal value as shown in Fig. \ref{topmass}.

\begin{figure}[h]
 \begin{center}
 \includegraphics[width=0.6\textwidth,natwidth=610,natheight=642]{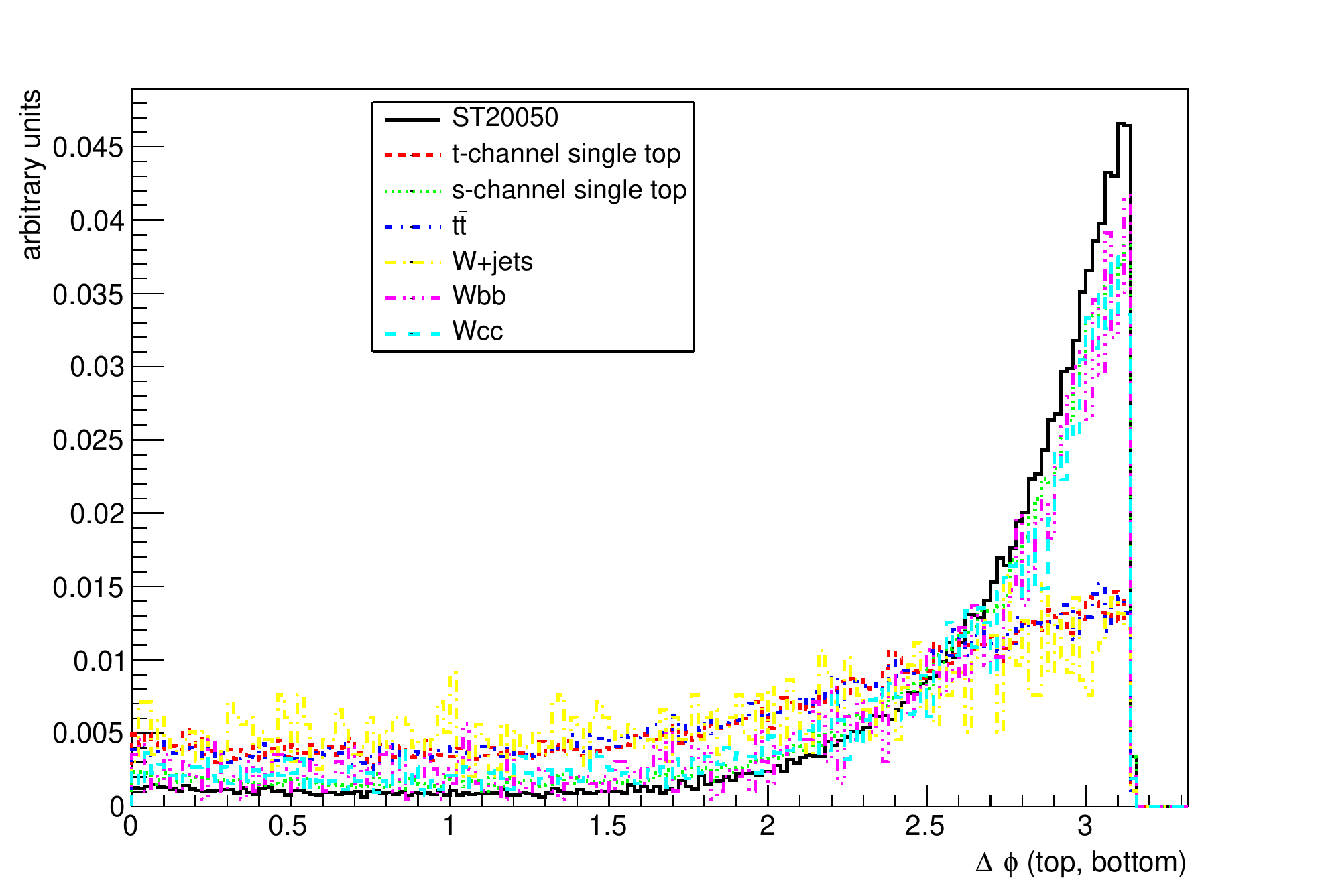}
 \end{center}
 \caption{The azimuthal angle difference is plotted between top and bottom quark to apply this cut in order to suppress those backgrounds which are not produce back-to-back.}
 \label{deltaphitb}
 \end{figure}

Furthermore, another important and interesting aspect of s-channel signal events is that they tend to produce the top and bottom quark pair in opposite directions due to the typical nature of s-channel processes. This feature should appear in azimuthal plane of the detector too. Therefore the azimuthal angle between the top quark and bottom quark is determined and the result is plotted as a distribution for both signal and background events for comparison as seen in Fig. \ref{deltaphitb}. According to the Fig.\ref{deltaphitb}, a selection cut is applied as $\Delta \phi$(top quark, bottom quark) $>$ 2.8. This cut will make the signal more visible on top of the background by increasing the signal to background ratio.

\begin{figure}[h]
 \begin{center}
 \includegraphics[width=0.6\textwidth,natwidth=610,natheight=642]{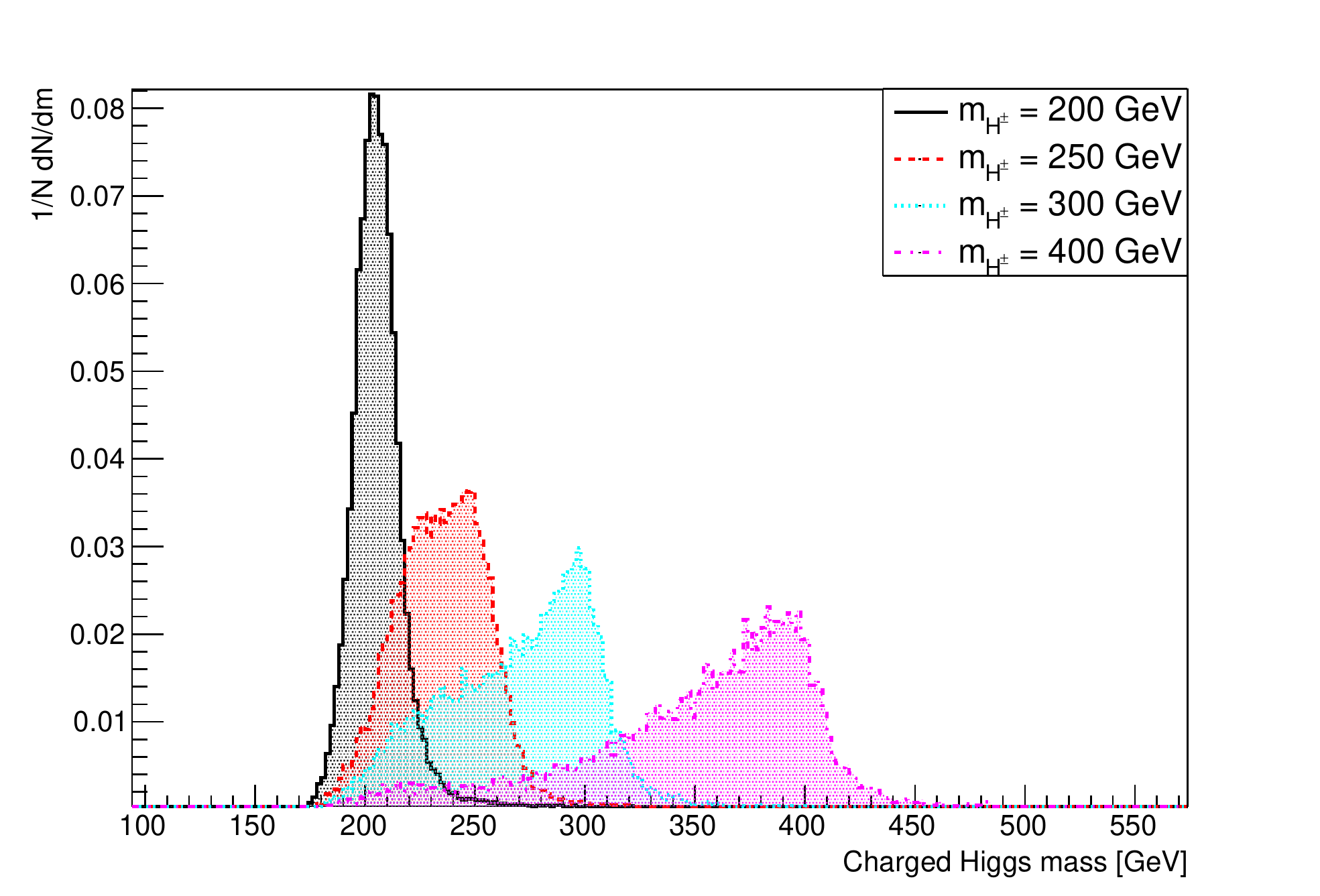}
 \end{center}
 \caption{The reconstructed charged Higgs mass distributions generated at different mass hypotheses.}
 \label{CHmassAll}
 \end{figure}

In signal events the top and bottom quark come from a charged Higgs boson and their invariant mass should in principle make the charged Higgs boson mass. However due to jet energy resolution, mis-identification of jets, errors in their energy and flight directions, and false jet combinations, a distribution of invariant mass with a peak at (almost) the nominal charged Higgs mass is obtained.  This distribution is seen in Fig.\ref{CHmassAll} where different charged Higgs mass hypotheses are tested in the simulation to make sure the obtained peak lies around the input mass. When all selection cuts are applied as in Tab. \ref{kincuts}, a chain of selection efficiencies is obtained for signal and background processes as shown in the Tab. \ref{signalcuts} and Tab.\ref{backgroundcuts}. The QCD multi jets sample is completely vanished even after generating millions of events with the limited computing resources.

\begin{table}[h]
\begin{tabular}{|c|c|}
\hline
Light jets & 2 highest $P_{T}$ jets, $E_{T}^{jets}>20$ GeV, $|\eta| < 3.0$ \\
\hline
b-jets with b-tagging & $E_{T}^{bjet} > 50$ GeV, $|\eta| < 3.0$  \\
\hline
W mass window & 60 GeV $<$ di-jet invariant mass ($m_{jj})< 100$ GeV \\
\hline
 Top quark mass window & 150 GeV $<$ jjb invariant mass ($m_{jjb}) < 190$ GeV \\
\hline
back-to-back production & $\Delta\phi$ (top,bottom) $>$ 2.8 \\
\hline
\end{tabular}
\caption{The selection cuts applied on both signal and background events. \label{kincuts}}
\end{table}

\begin{table}[h]
\begin{tabular}{|c|c|c|c|c|}
\hline
Selection Cut & Signal & Signal & Signal &Signal \\
& $M_{H^{\pm}}$ = 200 GeV & $M_{H^{\pm}}$ = 250 GeV & $M_{H^{\pm}}$ = 300 GeV & $M_{H^{\pm}}$ = 400 GeV \\
\hline
$\sigma$ x BR [pb]&5.63 & 4.68&2.73&0.98 \\
\hline
2 light jets &39.5$\%$ &41.8$\%$ &43.3$\%$ &46.5$\%$ \\
\hline
W mass cut &91.3$\%$ &91.6$\%$ &91.9$\%$ &91.8$\%$ \\
\hline
2 bjets &27.9$\%$ &15.9$\%$ &11$\%$ &7.3$\%$ \\
\hline
Top mass cut &63$\%$ &76.6$\%$ &68.4$\%$ & 56.3$\%$\\
\hline
$\Delta\phi$ cut &56.7$\%$ &50.9$\%$ &49$\%$ &54.4$\%$ \\
\hline
Total Efficiency &3.6$\%$ &2.4$\%$ &1.5$\%$ &0.96$\%$ \\
\hline
Expected events &20268 &11232 &4095 & 941\\
at 100 $fb^{-1}$ & & & & \\
\hline
\end{tabular}
\caption{Signal Efficiencies at different charged Higgs masses at tan$\beta$ = 50 are given. \label{signalcuts}}
\end{table}

\begin{table}[h]
\begin{tabular}{|c|c|c|c|c|c|c|c|}
\hline
Selection Cut & $t\bar{t}$ & SM single top & SM single top &Wjets& Wb$\bar{b}$ & Wc$\bar{c}$ & QCD \\
&  & s-channel & t-channel & & & &\\
\hline
$\sigma$ x BR[pb] & 285.4 & 5.8 & 133 & 1.69$\times10^{4}$ & 395 & 49 &1.169$\times 10^{8}$ \\
\hline
2 light jets &92.9$\%$ &40.5$\%$ &43.6$\%$ & 35$\%$& 32$\%$ & 36$\%$&0\\
\hline
W mass cut &31.9$\%$ &92$\%$ &93$\%$ &94$\%$& 96$\%$ &95$\%$ &0 \\
\hline
2 b-jets &11$\%$ &12$\%$ &9$\%$ &1.3$\%$ &6$\%$ &11$\%$ & 0\\
\hline
Top mass cut &60.9$\%$ &64.3$\%$ &61$\%$ &23$\%$ &13$\%$ &31$\%$ &0\\
\hline
$\Delta\phi$ cut &22.2$\%$ &48.8$\%$ &22.5$\%$ &19$\%$ &56$\%$ & 54$\%$&0\\
\hline
Total Efficiency &0.45$\%$ &1.42$\%$ &0.52$\%$ & 0.017$\%$ & 0.1$\%$ &0.6$\%$ & 0\\
\hline
Expected events & 128430 & 8236 &69160 & 287300 & 39500 & 29400 &0 \\
at 100 $fb^{-1}$ & & & & & & &\\
\hline
\end{tabular}
\caption{Selection efficiencies are shown for all background events. \label{backgroundcuts}}
\end{table}

The relative efficiencies are calculated for each selection cut with respect to previous cut when passing the signal and background samples through kinematic cuts. In this analysis a charged Higgs with $m_{H^{\pm}}$ = 180 GeV has not been considered because its mass is close to the top quark mass and is hard to observe due to a very limited phase space available for the charged Higgs decay to top and bottom quarks. This feature results in soft kinematics of the final state particles.\\

\begin{figure}[h]
 \begin{center}
 \includegraphics[width=0.6\textwidth,natwidth=610,natheight=642]{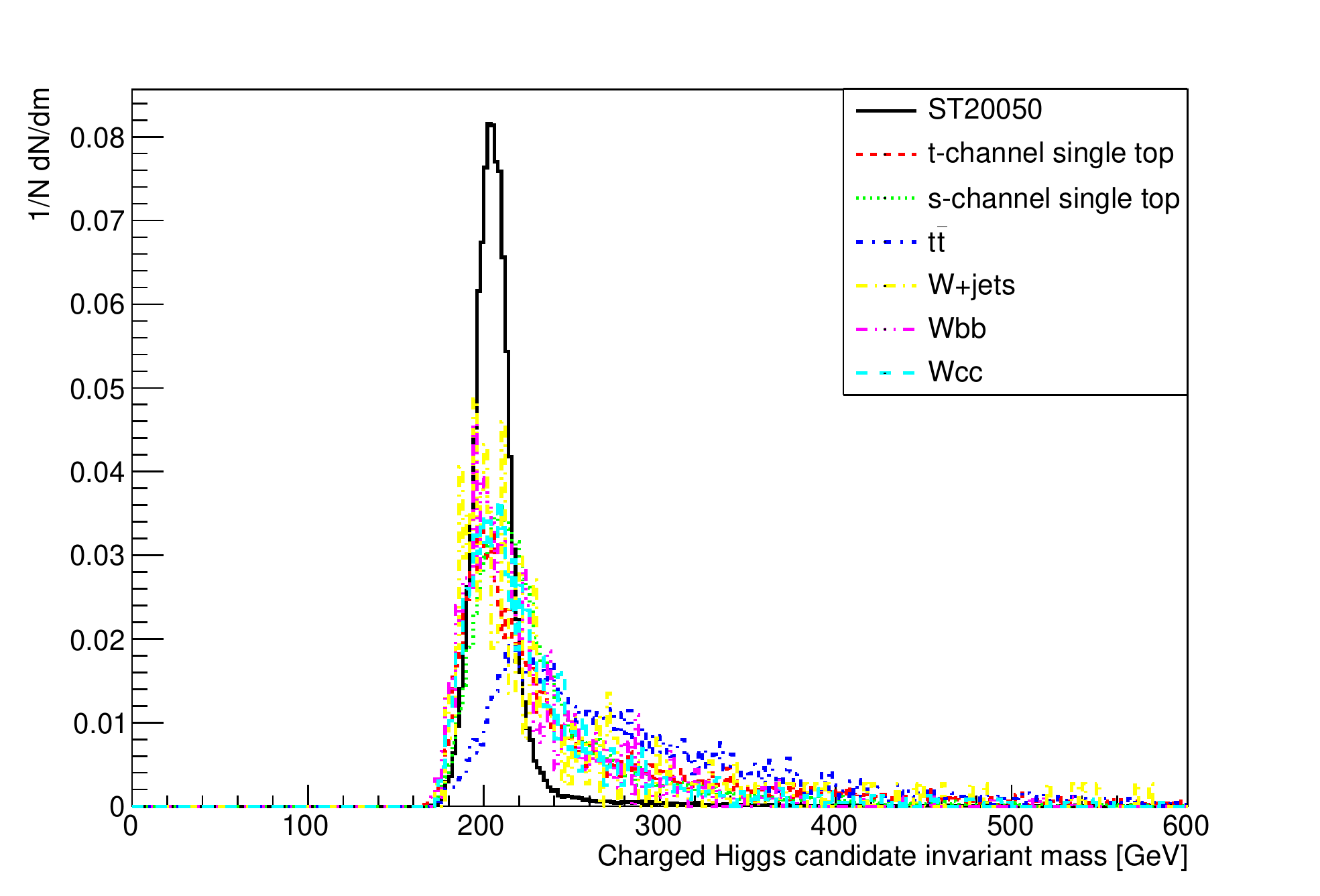}
 \end{center}
 \caption{The charged Higgs invariant mass distribution is obtained through 2 light jets and 2 b-jets combination. The signal  and background distributions are normalized to unity to find the most probable process over each other.}
 \label{CHmass}
 \end{figure}

\begin{figure}[h]
 \begin{center}
 \includegraphics[width=0.6\textwidth,natwidth=610,natheight=642]{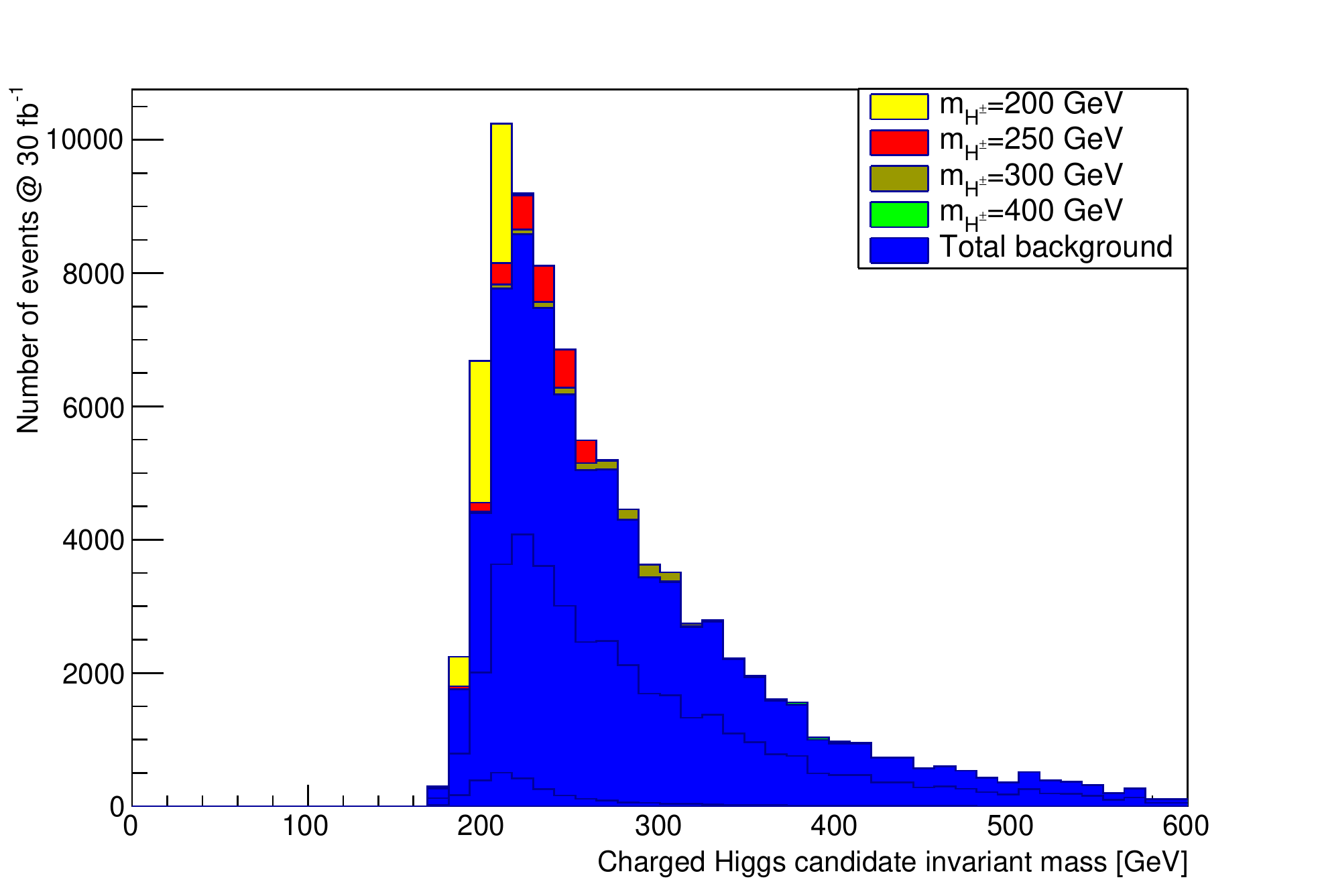}
 \end{center}
 \caption{Every sample is normalized to the real number of events obtained at 30 $fb^{-1}$}. The charged Higgs is on the top of the total background events at different charged Higgs mass hypotheses independently at tan$\beta = 50$. It shows its visibility for charged Higgs observability. Only dominant backgrounds are labeled.
 \label{massbin}
 \end{figure}

The charged Higgs is reconstructed through $jjjb$ combination which is considered as the charged Higgs candidate as plotted in Fig. \ref{CHmass} with all dominant backgrounds. Charged Higgs "ST20050" mass peak can be seen clearly with significant background suppression. However, there are always fraction of fake entries from background and systematic uncertainties. If each distribution is normalized to the real number of events at 30 $fb^{-1}$ including selection efficiencies, Fig.\ref{massbin} is obtained.\\
In the hadron collider experiments, the realistic approach needs to take care all the sources of uncertainties which must be taken into account including electronic noise, pile-up, trigger, vertex etc. To assess the impact of systematic uncertainties arising from detector simulation, the selection cuts are re-applied after shifting a particular parameter up and down by one unit of uncertainty. In this analysis, where the final state is fully hadronic and large number of jets are expected, the jet energy uncertainty is expected to be the dominant source of uncertainty at the current LHC stage and it may be less than 1$\%$ in the central part of the detector for jets having transverse energies in the range 55 to 500 GeV \cite{syserror}. The correction coefficient  of the jets four momentum may include several multiplicative factors for Data/MC calibration, jet energy scale uncertainties and off-set effects. In addition some other sources of uncertainties are also expected e.g., the uncertainty from the fit function, the uncertainty on the b-tagging (mis-)identification efficiency and the background modeling contributing in the total background probability density function. The latter part essentially relies on the correct understanding of background distributions which is well achievable in the real data analysis where the distributions of the different backgrounds are taken from real data and then MC are used for comparison to obtain a reasonable parton density function of the total background. The uncertainties on the scale factors arise from the statistical uncertainty of the factors; the effect of binning in trigger periods, the effect of binning in number of tracks associated with the selected jets and a potential kinematic bias, evaluated by varying the jet $p_{T}$ selection criterion. These systematic uncertainties are strongly correlated by statistical effects, so they are each considered individually in the complete analysis, rather than as a combined uncertainty. So the detailed calculation of systematics is beyond the scope of this analysis.\\

Finally in Tab. \ref{Significance} the number of signal and background events, corresponding efficiencies to the charged Higgs mass windows, S/B ratio and the optimized signal significance ($S/\sqrt{B}$) are shown. The S/B approaches to its best value around 37$\%$. The charged Higgs mass window is applied in the specific region where a maximum signal significance is achieved. This condition suppress significant amount of background events. The QCD jets are restricted at the jets-quark matching stage. By generating a large statistics of QCD sample not even few events could survived effectively.  \\

\begin{figure}[h]
 \begin{center}
 \includegraphics[width=0.6\textwidth,natwidth=610,natheight=642]{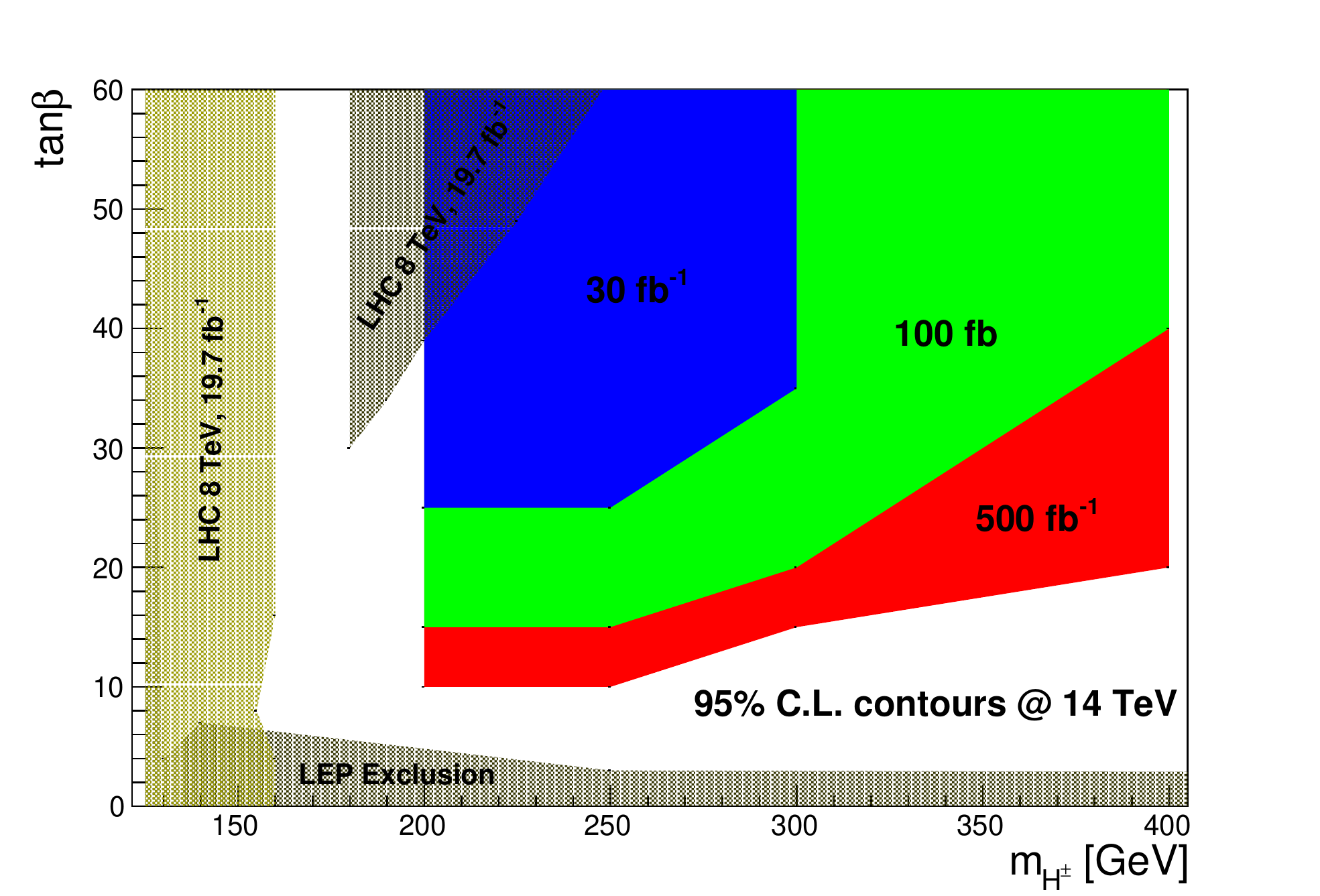}
 \end{center}
 \caption{The exclusion curves shown with red, blue and green color bands corresponds to 95$\%$ $C.L.$ contour as a function of charged Higgs mass and tan$\beta$ at three different integrated luminosities.}
 \label{95CL}
 \end{figure}

\begin{figure}[h]
 \begin{center}
 \includegraphics[width=0.6\textwidth,natwidth=610,natheight=642]{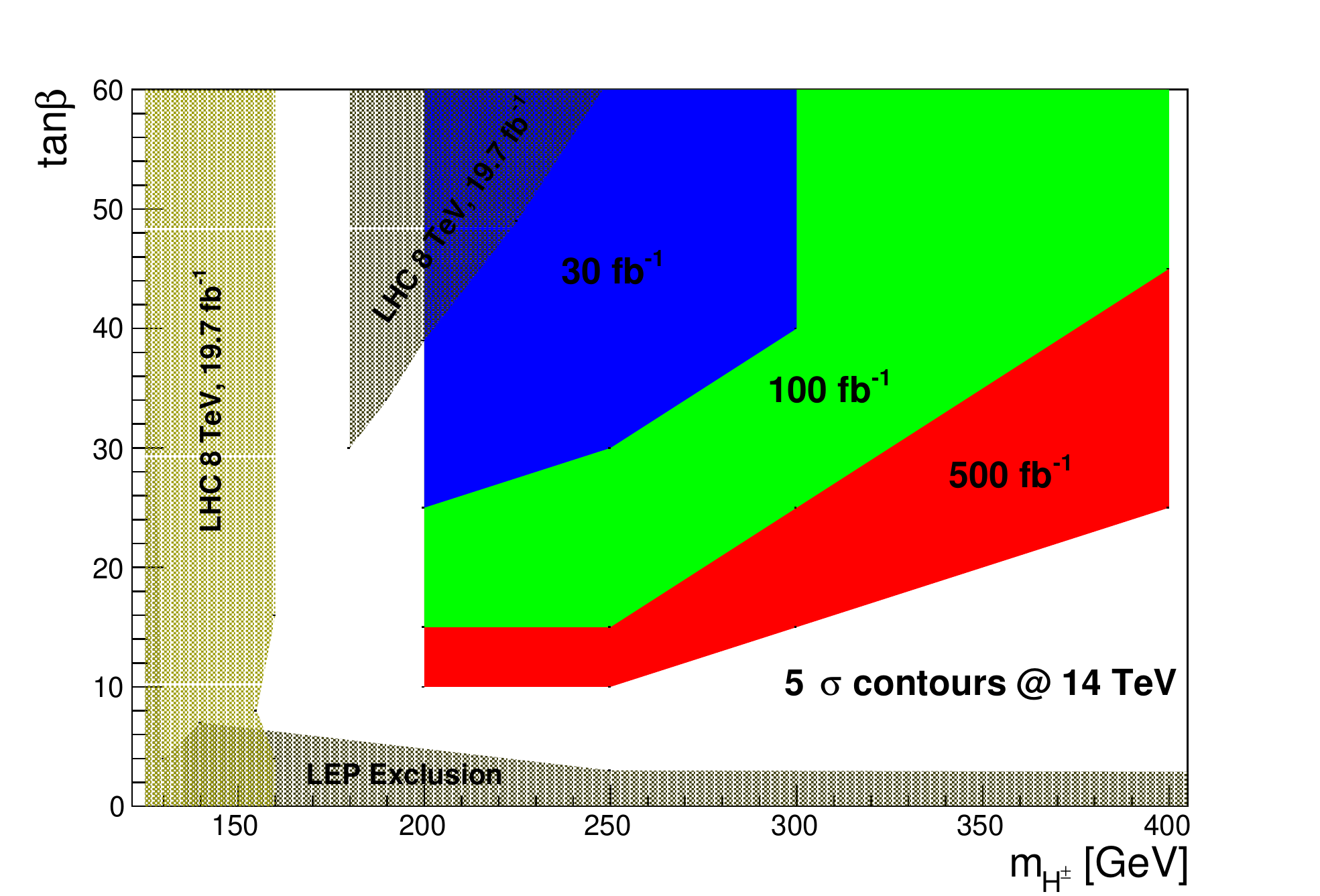}
 \end{center}
 \caption{The 5 $\sigma$ discovery contours are plotted here as a function of charged Higgs at various integrated luminosities.}
 \label{5sigma}
 \end{figure}

At the end to demonstrate the results validity in the MSSM parameter space within the presence of all previous experimental constraints, 5$\sigma$ discovery contours and exclusion curves at 95$\%$ Confidence Level (C.L.) are obtained by scanning the chosen charged Higgs mass points and tan$\beta$ values. To perform this algorithm, the TLimit class implemented in ROOT \cite{root} is used to obtain both contours. The results are shown in Figs. \ref{95CL} and \ref{5sigma} together with the previously excluded areas by LEP and LHC 8 TeV data. As is seen, a wide range of parameter space is still available for the discovery of charged Higgs.

\section{Conclusions}
The s-channel single top process was studied as a source of charged Higgs in the fully hadronic final state at LHC. Kinematic selection cuts were designed to increase the signal to background ratio and signal significance at 14 TeV center of mass energy. On the basis of exclusion contours at 95$\%$ C.L. and 5$\sigma$ discovery contours, it was shown that the charged Higgs signal can be well observed or excluded in a wide range of ($m_{H^{\pm}}$,tan$\beta$) phase space. This process requires large tan$\beta$ $>$ 25 at 30 $fb^{-1}$ integrated luminosity and can probe the area up to tan$\beta$ $>$ 10 at 500 $fb^{-1}$. In all the above calculations the systematic and theoretical uncertainties are not taken into account. However, comparing results presented in this analysis and previous simulations performed at CMS and ATLAS experiments and the current LHC results, the channel proposed in this work can be considered as a complementary channel to other search channels and help increasing the signal statistics in case of the charged Higgs existence.

\begin{table}[h]
\begin{tabular}{|c|c|c|c|c|c|c|}
\hline
Sample &\multicolumn{2}{c}{Mass window} & Total & no. of & S/B & Optimized  \\
 & Lower limit & Upper limit & Efficiency & Events & &S/$\sqrt{B}$ \\
\hline
Signal, $m_{H^{\pm}}$ = 200 GeV & 192 & 216 & 0.031 & 5244 & 0.3767 & 44\\
 Total Background & 192 & 216 &  & 13920 & & \\
\hline
Signal, $m_{H^{\pm}}$ = 250 GeV & 228 & 264 & 0.01679 & 2356 & 0.0864 & 14\\
Total Background & 228 & 264 &  & 27267 & & \\
\hline
Signal, $m_{H^{\pm}}$ = 300 GeV& 276 & 324 & 0.00846 & 692 & 0.0368 & 5  \\
 Total Background & 276 & 324 &  & 18827 &  & \\
\hline
Signal, $m_{H^{\pm}}$ = 400 GeV& 396 & 564 & NS  & 93 & 0.0108 & 1\\
Total Background & 396 & 564 &  &  8600 &  &  \\
\hline
\end{tabular}
\caption{Signal to background ratio and signal significance values obtained for four different samples, where "NS" represents negligibly small. \label{Significance}}
\end{table}

\end{document}